\documentclass[aps,prl,reprint]{revtex4-1}

\usepackage{graphicx}
\usepackage{amsmath}
\usepackage{amssymb}

\newcommand\rpict[1]{\ref{fig:#1}}
\newcommand\leqt[1]{\protect\label{eq:#1}}
\newcommand\reqtn[1]{\ref{eq:#1}}
\newcommand\reqt[1]{(\reqtn{#1})}

\begin{document}

\title{Modulated coupled nanowires for ultrashort pulses}

\author{Alexander S. Solntsev}

%\email{Alexander.Solntsev@anu.edu.au}

\affiliation{Nonlinear Physics Centre,
% and Center for Ultrahigh-bandwidth Devices for Optical Systems,
Research School of Physics and Engineering, Australian National University, Canberra
ACT 2601, Australia}

\author{Andrey A. Sukhorukov}

\affiliation{Nonlinear Physics Centre,
% and Center for Ultrahigh-bandwidth Devices for Optical Systems,
Research School of Physics and Engineering, Australian National University, Canberra
ACT 2601, Australia}
%\\
%$^\ast$Corresponding author: Alexander.Solntsev@anu.edu.au}

%\dates{Compiled \today}

%\ociscodes{080.1238 (Array waveguide devices); 320.5540 (Pulse shaping).}

%\doi{\url{http://dx.doi.org/10.1364/ao.XX.XXXXXX}}

\begin{abstract}
We predict analytically and confirm with numerical simulations that inter-mode dispersion in nanowire waveguide arrays can be tailored through periodic waveguide bending, facilitating flexible spatio-temporal reshaping without break-up of femtosecond pulses. This approach allows simultaneous and independent control of temporal dispersion and spatial diffraction that are often strongly connected in nanophotonic structures.
\end{abstract}

%\setboolean{displaycopyright}{true}

\maketitle
%\thispagestyle{fancy}
%\ifthenelse{\boolean{shortarticle}}{\abscontent}{}

High-index-contrast nanowires offer unique advantages for manipulation of optical pulses in compact photonic circuits, providing high field confinement and enabling precise dispersion engineering~\cite{Turner:2006-4357:OE, Foster:2008-1300:OE}. In particular optical chips based on silicon sub-wavelength waveguides allow for efficient frequency
conversion~\cite{Foster:2006-960:NAT, Solntsev:2012-446:OL}, all-optical pulse control~\cite{Almeida:2004-1081:NAT} and all-optical switching~\cite{Vlasov:2008-242:NPHOT}. Furthermore, couplers~\cite{deNobriga:2010-3925:OL} and arrays of coupled nanowire waveguides~\cite{Peleg:2009-163902:PRL, Benton:2009-5879:OE} open possibilities for efficient spatio-temporal shaping of optical pulses. In order to harness these opportunities, it is essential to develop approaches to simultaneously and independently control temporal and spatial dispersion, as these characteristics can be strongly connected in nanophotonic structures.
This connectivity can lead to difficulties designing a waveguide array supporting propagation of ultrashort pulses, since pulses either disperse due to strong temporal dispersion or break-up due to strong spatial diffraction~\cite{Benton:2009-5879:OE}.

One possible approach to achieve required spatio-temporal dispersion is to carefully design waveguide array geometry and use complex photonic crystal structures~\cite{Laegsgaard:2004-2473:OL}. That is however a very complicated method. Another way to tailor dispersion is by introducing periodic waveguide bending~\cite{Garanovich:2007-475:OL, DellaValle:2010-673:OL, Garanovich:2012-1:PRP}. This approach allows relatively simple fabrication and offers substantial design flexibility. Periodic waveguide bending was introduced as an effective tool for polychromatic diffraction management~\cite{Garanovich:2012-1:PRP, Garanovich:2007-475:OL, Szameit:2009-271:NPHYS}, however it has only been studied in the context of continuous light illumination and conventional micro-scale waveguides.
In this work, we develop an approach to simultaneously control spatial and temporal dispersion and demonstrate through numerical simulations the application of this concept to the suppression of ultrashort pulse distortion and break-up in nano-waveguide arrays.
%we develop a model that incorporates periodic waveguide array bending and ultrashort pulse dynamics, and demonstrate through numerical simulations the application of this concept to suppress pulse break-up and pulse dispersion in nano-waveguide arrays.

Ultrashort pulses have a broad spectrum encompassing a large range of frequencies $\omega$, therefore temporal dispersion characterized by a propagation constant $\beta_s(\omega)$ has a significant influence on the pulse dynamics. In waveguide arrays pulses can also switch between different waveguides. One of the most important parameters characterizing waveguide arrays is a coupling coefficient $C_s(\omega)$, which determines the rate at which light couples between the neighboring waveguides and thus regulates the spatial dispersion. The coupling coefficient for straight lossless waveguides is real: $C_s(\omega)={\rm Re}[C_s(\omega)]$.

It was shown previously~\cite{Garanovich:2012-1:PRP} that bending the waveguides affects the coupling by introducing an additional phase shift.
%$\phi$: $C(\lambda)={\rm Re}[C(\lambda)] \exp \{\imath \phi\}$ .  
If all waveguides in an array have the same
%periodically modulated 
bending profile $x_0(z)$,
% \equiv x_0(z+L_b)$, 
where $x_0$ is a transverse coordinate of the waveguide center and $z$ is the propagation direction,
% and $L_b$ is a modulation period, 
then the complex electric field amplitude $E_n$ in $n$-th waveguide of the  array satisfies the following coupled-mode equations~\cite{Garanovich:2012-1:PRP}:
\begin{equation} \leqt{frequency_domain}
\begin{split}
   \imath \partial_z E_n & (z, \omega) + 
   \beta_s(\omega) E_n(z, \omega) = \\
   & -C_s(\omega) \exp{[\imath n_0 d_w \dot{x}_0(z) \omega / c]} E_{n-1}(z, \omega)  \\
   & - C_s(\omega) \exp{[-\imath n_0 d_w \dot{x}_0(z) \omega / c]} E_{n+1}(z, \omega).
\end{split}
\end{equation}
Here $n_0$ is an effective refractive index, $\omega = 2 \pi c / \lambda$ is the angular frequency, $\lambda$ is a light wavelength in vacuum, and $d_w$ is the distance between the coupled waveguides.
%, and $\beta_s$ is the propagation constant in a waveguide.

We extend this method to consider the dynamics of ultrashort pulses and study its applicability to nanophotonic structures. In low-index waveguide arrays both the propagation constant of individual waveguides and the coefficient characterizing coupling between the neighboring waveguides are mildly dispersive. In contrast, in nanowire high-index waveguides the dispersion can be much stronger, and also small changes of waveguide cross-section dramatically affect both temporal dispersion and spatial diffraction~\cite{Benton:2009-5879:OE}.

To investigate the pulse dynamics in the coupled nanowires we combine the approaches previously developed for the description of nanowire arrays~\cite{Benton:2009-5879:OE} and curved conventional waveguide arrays~\cite{Garanovich:2007-475:OL, Szameit:2009-271:NPHYS}. We derive the following system of equations by applying to Eq.~\reqt{frequency_domain} the Fourier transform $E_n(z,t)=\int {\rm d} \omega E_n(z,\omega) \exp{(-\imath \omega t)}$, where $t$ is time, and perform a Taylor expansion of coupling and propagation coefficients:
% up to a second order:
%
\begin{equation} \leqt{main}
\begin{split}
   i \partial_z E_n (z, t) &
%   + \imath \beta_2 \partial_t^2 E_n (z, t) / 2 = \\
   + \hat{\beta} E_n (z, t) = \\ 
   & - \hat{C}(z) E_{n-1} (z, t) - \hat{C}^\ast(z) E_{n+1} (z, t).
\end{split}
\end{equation}
Here $\hat{\beta}$ determines the temporal dispersion in a waveguide, and $\hat{C}$ characterizes the coupling between the neighboring waveguides:
%$\beta_2$ is the group-velocity dispersion coefficient, and $\hat{C}_{+,-} (\imath \partial_t)$ is a Taylor expansion of the coupling coefficient:
%
\begin{equation} \leqt{coupling}
   \hat{\beta} = \sum_{m=0}^M \frac{ \beta_m }{m!} (\imath \partial_t)^m , \,
   \hat{C}(z)  =    \sum^M_{m=0} \frac{ c_m(z) }{m!} (\imath \partial_t)^m ,
\end{equation}
where $M$ is a sufficiently large number to capture the dispersion features over the pulse bandwidth. The Taylor coefficients are
%$c_m^{(+,-)}$ are the Taylor coefficients:
%
\begin{equation} \leqt{Taylor}
\begin{split}
  \beta_m &= (\partial_\omega)^m \beta_s(\omega)|_{\omega=\omega_0}, \\
  c_m(z) &= (\partial_\omega)^m \left[C_s(\omega) \exp{(\imath n_0 d_w \dot{x}_0(z) \omega / c)}\right]_{\omega=\omega_0},
\end{split}
%c_m^{(+)} = c_m^0 \exp\left[ \imath \phi(z) \right],\\ \nonumber
%c_m^{(-)} = c_m^0 \exp\left[- \imath \phi(z) \right].
\end{equation}
where $\omega_0$ is the central pulse frequency.
Note that the dispersion coefficients ($\beta_m$) do not depend on the coupling. On the other hand, the coupling coefficients ($c_m$) depend non-trivially on the propagation distance ($z$) through an interplay between the dispersion of coupling between the straight waveguides ($C_s$) and the bending profile [${x}_0(z)$] induced dispersion.
%They incorporate the coupling dispersion of the straight waveguides $c_m^0$ and the phase modulation due to possible periodic waveguide bending $\exp[\pm \imath \phi(z)]$. The phase $\phi(z)$ has the following form:
%
%\begin{equation} \leqt{phi}
%\phi(z)=-2 \pi n_0 d_w \dot{x}_0(z) / \lambda_0.
%\end{equation}

%--------------------------------------------------------------------------------------------
\begin{figure}[t]
\centering
{\includegraphics[width=\linewidth]{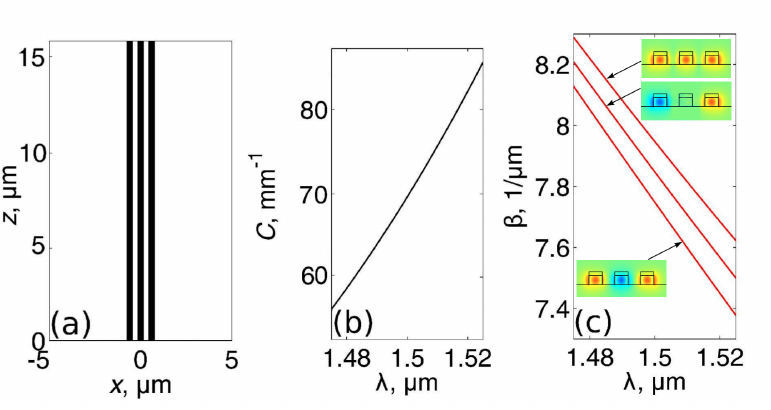}}
\caption{
(a)~The scheme of three straight coupled nanowires.
(b)~Coupling coefficient vs. wavelength between the neighboring waveguides.
(c)~Supermode propagation constants vs. wavelength (solid lines), and the corresponding profiles of the supermodes (cross sections of three coupled waveguides).}
\label{fig:modesStraight}
\end{figure}
%--------------------------------------------------------------------------------------------

It was shown~\cite{Garanovich:2012-1:PRP} that for polychromatic light propagation in periodically curved waveguide arrays, after 
the full bending period the beam diffraction is the same as in a straight array with the effective coupling coefficient. We check that the same approach can be applied to the pulse propagation when $x_0(z) \equiv x_0(z+L_b)$ and $L_b$ is a modulation period, and the effective coupling is:
%For curved wavegid
%Similar to the previous analysis Ref.~\cite{Garanovich:2012-1:PRP}, we calculate
%  that after the full bending period ($z \rightarrow z + L_b$) the beam diffraction in the periodically
%curved waveguide array is approximately the same as in a straight array with the effective coupling coefficient:
%
\begin{equation} \leqt{effective}
   C^{\rm eff}(\omega) = C_s(\omega) L_b^{-1} \int_0^{L_b} \cos[n_0 d_w \dot{x}_0(z) \omega / c] \mathrm{d} z .
\end{equation}
The corresponding Taylor expansion coefficients are:
\begin{equation} \leqt{TaylorEffective}
  c_m^{\rm eff} = (\partial_\omega)^m C^{\rm eff}(\omega)|_{\omega=\omega_0}.
\end{equation}
We see that diffraction of beams is defined by an interplay of the additional bending-induced dispersion introduced through the frequency dependence of the integral in Eq.~\reqt{effective}, and the intrinsic frequency dependence of the coupling coefficient in a straight waveguide array $C_s(\omega)$.

We investigate the influence of the periodic waveguide bending on the pulse reshaping, and consider a representative example of 
%. We check several types of periodic bending profiles, and choose 
a cosine profile with the amplitude $A$ and period $L_b$:
\begin{equation} \leqt{cos_profile}
   x_0(z)=A \cos(2 \pi z / L_b ).
\end{equation}
Below we show that a special combination of the modulation parameters allows us to suppress the dispersion of the effective coupling coefficient and accordingly avoid the pulse distortion.
%coupling coefficient dispersion and at the same time have low propagation dispersion.

To demonstrate our approach, we consider coupled Si nanowire waveguides and use COMSOL RF module to calculate the dispersion in straight coupled waveguides. The dimensions of the wires are as follows. The wires are 220\,nm high and 330\,nm wide, placed on a silica slab. There is a 100\,nm high etching mask with refractive index 1.35 on top of wires. Otherwise the wires are surrounded by air. We choose these parameters to obtain nearly zero group velocity dispersion $\beta_2 \approx 0$ in the proximity of $\lambda_0 =1.5\,\mu \mathrm{m}$ wavelength for a single nanowire, as this would minimize the pulse distortion.
We however emphasize that even in this regime, a pulse can exhibit distortion due to the dispersion of coupling between the nanowires in an array~\cite{Benton:2009-5879:OE}.

%--------------------------------------------------------------------------------------------
\begin{figure}[t]
\centering
{\includegraphics[width=\linewidth]{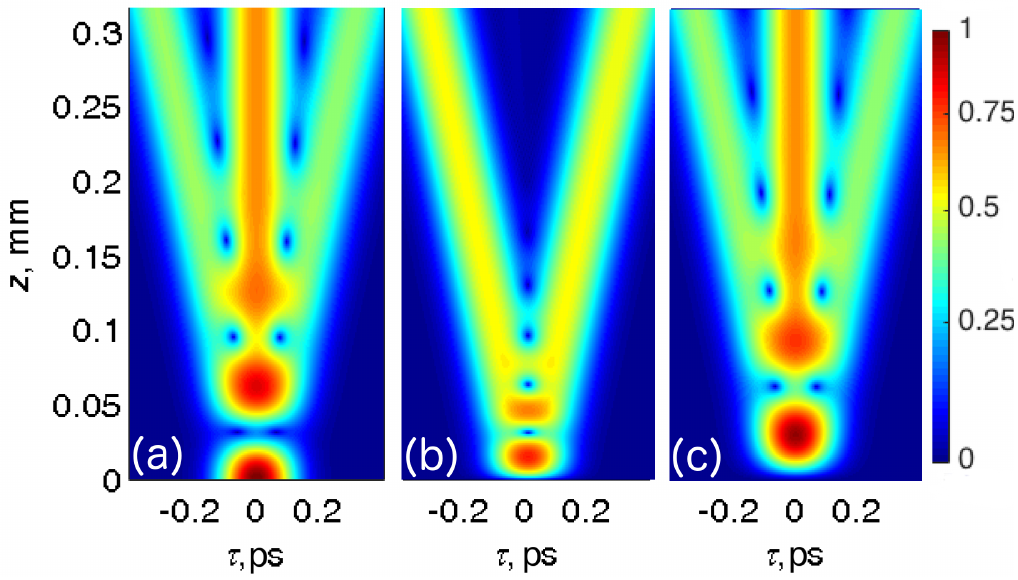}}
\caption{Pulse intensity evolution along straight coupled nanowires: (a)~left, (b)~central, and (c)~right nanowire.
The temporal axis corresponds to a moving time frame with the group velocity at the central wavelength, $\tau = t - z \beta_1$.}
\label{fig:intensityStraight}
\end{figure}
%--------------------------------------------------------------------------------------------

To determine the coupling strength between the neighboring waveguides, we follow the approach of Ref.~\cite{Benton:2009-5879:OE} and analyze a two-waveguide coupler with wire-to-wire separation of 330\,nm. We calculate the propagation constants for symmetric and antisymmetric supermodes of the coupler $\beta^{\rm sym}(\omega)$ and $\beta^{\rm asym}(\omega)$, respectively. The propagation constant for a single waveguide can be well approximated by the average of the symmetric and antisymmetric supermode propagation constants, $\beta_s(\omega) \approx [\beta^{\rm sym}(\omega)+\beta^{\rm asym}(\omega)]/2$, while the difference defines the coupling coefficient $C_s(\omega) = [\beta^{\rm sym}(\omega)-\beta^{\rm asym}(\omega)]/2$.  The coefficients of the Taylor expansion of the propagation constant are found as $\beta_0 = 7.85 \, \mu{\rm  m^{-1}}$, $\beta_1 = 17 \, {\rm fs} \, \mu {\rm m^{-1}}$, $\beta_2 = 1.4 \, {\rm fs^2 \,} \mu{\rm m^{-1}}$. Accordingly, we find that the group velocity dispersion is indeed suppressed for pulses with duration down to 100\,fs.
%We also calculate the second order of dispersion for the propagation constant $\beta_m=\partial_{\omega}^m [\beta_s(\omega)+\beta_a(\omega)]/2$ and the coupling coefficient $c_m^{(0)}=\partial_{\omega}^m |\beta_s(\omega)-\beta_a(\omega)|/2$ in the vicinity of $\lambda_0 =1.5\, \mathrm{\mu m}$.

Next we consider an array of three coupled nanowires with all parameters as noted before, see Fig.~\rpict{modesStraight}(a). We use a three-waveguide system for all the following calculations, since it allows us to fully demonstrate the pulse control method proposed in this work. In agreement with predictions of Ref.~\cite{Benton:2009-5879:OE}, we notice that the variations of the coupling coefficient for such waveguide array are significant across a relatively narrow spectrum [see Fig.~\rpict{modesStraight}(b)], which could lead to temporal reshaping of short pulses during propagation. The Taylor expansion of the coupling coefficient for straight waveguides is $c_0 = 69.7 \, {\rm mm^{-1}}$, $c_1 = -0.71 \, {\rm fs \, }\mu{\rm m^{-1}}$, $c_2 = 3.6 \, {\rm fs^2 \, }\mu{\rm m^{-1}}$, which reveals strong linear dispersion and small quadratic dispersion in the wavelength range between 1.46\,$\mu{\rm m}$ and 1.54\,$\mu{\rm m}$.
The coupled waveguides support three supermodes, which characteristic spatial profiles are shown in Fig.~\rpict{modesStraight}(c). We calculate the dependence of the supermode propagation constants on the wavelength, see Fig.~\rpict{modesStraight}(c). These dependencies have different slopes corresponding to different supermode velocities. We show below that this leads to pulse splitting, which can be suppressed via periodic waveguide bending.

%--------------------------------------------------------------------------------------------
\begin{figure}[t]
\centering
{\includegraphics[width=\linewidth]{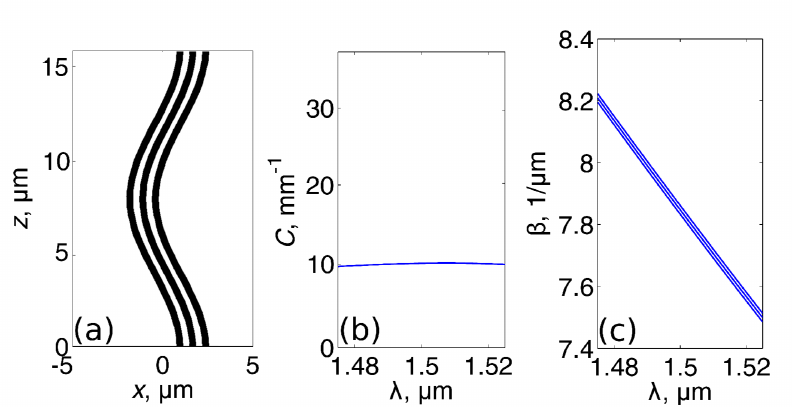}}
\caption{(a)~The scheme of three periodically bent coupled nanowires with the bending period $L_b = L_C = 15.75\,mu {\rm m}$.
(b)~Effective coupling coefficient over one bending period between the neighboring nanowires vs. wavelength.
(c)~Average supermode propagation constants over one bending period vs. wavelength.}
\label{fig:modesCurved}
\end{figure}
%--------------------------------------------------------------------------------------------

First we analyze the pulse dynamics in straight waveguides with length $L = 20 L_C = 315\,\mu{\rm m}$, where $L_C=15.75\,\mu{\rm m}$ is the length required for full coupling from one waveguide to another at the wavelength $\lambda_0$. As an input, we consider a single 100\,fs long transform-limited Gaussian pulse with the central wavelength $\lambda_0$ coupled to the left nanowire of the straight waveguide array. Figs.~\rpict{intensityStraight}(a-c) demonstrate that initially the pulse couples from the left (a) to the central (b) and then to right (c) nanowire without significant distortions. Then the pulse starts to split into three separate pulses in the edge waveguides (a,c) and into two pulses in the central waveguide (b), in agreement with the previous study~\cite{Benton:2009-5879:OE}. These pulses propagate with different group velocities, which correspond to three different supermode velocities supported in the structure [see Fig.~\rpict{modesStraight}(c)]. Such behavior demonstrates that although the single nanowire propagation dispersion can be engineered, spatial diffraction in arrays of nanowires is still strongly limited by the coupling dispersion. Moreover the propagation dispersion and the coupling dispersion in nanowire waveguide arrays are interconnected, and therefore an approach allowing for the independent control of these characteristics would offer essential benefits for various applications.

Next, we investigate the influence of the periodic waveguide bending on the pulse reshaping. We choose a bending profile according to Eq.~\reqt{cos_profile}. 
%We show that it allows to considerably flatten the effective coupling coefficient dispersion around $\lambda_0 =1.5\,\mathrm{\mu m}$ for a certain bending amplitude $A$.
We vary the bending amplitude $A$ and search for the minima of the coupling dispersion $\partial C_{eff} / \partial \omega$ in the vicinity of $\lambda_0$.  We choose the bending period $L_b = L_C = 15.75\,\mu{\rm m}$, as it allows us to consider nanowires with smaller curvature for the purposes of easier potential fabrication and reduction of propagation losses. As we show bellow, one can choose a bending profile that simultaneously allows for a strong coupling dispersion control and does not introduce bending propagation losses.

%--------------------------------------------------------------------------------------------
\begin{figure}[t]
\centering
{\includegraphics[width=\linewidth]{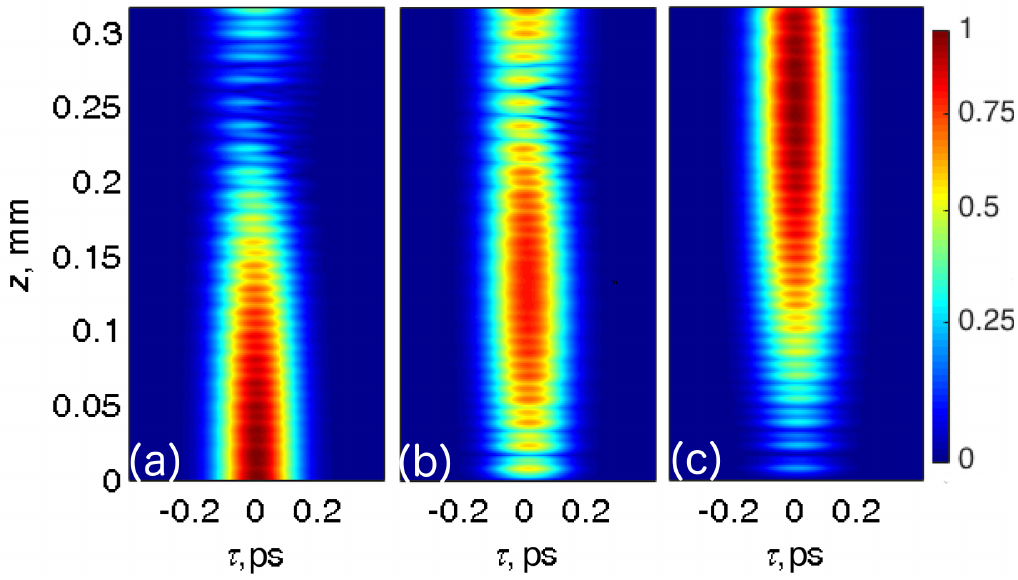}}
\caption{Pulse intensity evolution along periodically curved coupled nanowires: left edge nanowire (a), central nanowire (b), right edge nanowire (c). The temporal axis corresponds to a moving time frame with the group velocity at the central wavelength, $\tau = t - z \beta_1$.}
\label{fig:intensityCurved}
\end{figure}
%--------------------------------------------------------------------------------------------

%
We calculate the effective coupling coefficient $C_{eff}$ using Eq.~\reqt{effective}. We choose the value of $A=A_{min}$ corresponding to the first minimum of $\partial C_{eff} / \partial \omega$, and accordingly the smallest suitable bending curvature. The optimal bending amplitude is found to be equal to $A_{min} = 1.3\,\mu{\rm m}$. The bending losses for the corresponding curvature value should be practically absent according to the previous studies of bent nanowire waveguides~\cite{Dai:2007-2853:JOSB}.
The resulting effective coupling coefficient $C_{eff}$ shown in Fig.~\rpict{modesCurved}(b) becomes almost constant over a broad spectral region in comparison to that for the straight waveguides [see Fig.~\rpict{modesStraight}(b)]. The Taylor expansion of the effective coupling coefficient is $c_0^{\rm eff} = 9.96 \, {\rm mm^{-1}}$, $c_1^{\rm eff} = -0.0076 \, {\rm fs \, }\mu{\rm m^{-1}}$, $c_2^{\rm eff} = -14.2 \, {\rm fs^2 \, }\mu{\rm m^{-1}}$. Although the quadratic coupling dispersion is slightly increased in comparison to the straight waveguide array, the linear coupling dispersion, which is the main temporal reshaping driver for 100\,fs long pulses in such structures, is suppressed by two orders of magnitude compared to the straight waveguides.
%{\bf specify the values of $c_m^{\rm eff}$ coefficients for $m=0,1,2,3,?$}
In Fig.~\rpict{modesCurved}(c) we plot the supermode propagation constants for the curved waveguide arrays with three coupled nanowires calculated with the use of the effective coupling coefficient. The propagation constants for the three supermodes now have similar slopes, which suggests that short-pulse break-up due to coupling would be suppressed.

We now calculate the intensity evolution of a 100\,fs transform-limited pulse coupled to the left nanowire of the periodically curved waveguide array using the Eqs.~\reqt{main}-\reqt{Taylor}. We see in Figs.~\rpict{intensityCurved}(a-c) that as a result of vanishing coupling dispersion the temporal pulse break-up is suppressed, and a pulse can now be switched as a whole between the waveguides. Thus the temporal and the spatial dispersion in nanowire waveguide arrays can be controlled independently via single waveguide dispersion engineering and periodic waveguide bending.

These results demonstrate that spatio-temporal dispersion engineering in high-index-contrast nanowire waveguide arrays can be efficiently realized through the introduction of periodic waveguide bending, which can enable flexible spatio-temporal manipulation of femtosecond pulses. We anticipate that these results will open novel approaches to on-chip all-optical light control~\cite{Peleg:2009-163902:PRL}. This approach can also be useful for enhanced parametric frequency conversion~\cite{Foster:2008-1300:OE} and broadband photon-pair generation and quantum walks~\cite{Solntsev:2012-27441:OE}.

The work was supported by NCI National Facility and the Australian Research Council, including Discovery Project DP130100086 and Future Fellowship FT100100160.
We acknowledge useful discussions with Ivan Garanovich.


\begin{thebibliography}{20}
\newcommand{\enquote}[1]{``#1''}

\bibitem{Turner:2006-4357:OE}
A.~C. Turner, C.~Manolatou, B.~S. Schmidt, M.~Lipson, M.~A. Foster, J.~E.
  Sharping, and A.~L. Gaeta, \enquote{Tailored anomalous group-velocity
  dispersion in silicon channel waveguides,} Opt. Express \textbf{14},
  4357--4362 (2006).

\bibitem{Foster:2008-1300:OE}
M.~A. Foster, A.~C. Turner, M.~Lipson, and A.~L. Gaeta, \enquote{Nonlinear
  optics in photonic nanowires,} Opt. Express \textbf{16}, 1300--1320 (2008).

\bibitem{Foster:2006-960:NAT}
M.~A. Foster, A.~C. Turner, J.~E. Sharping, B.~S. Schmidt, M.~Lipson, and A.~L.
  Gaeta, \enquote{Broad-band optical parametric gain on a silicon photonic
  chip,} Nature \textbf{441}, 960--963 (2006).

\bibitem{Solntsev:2012-446:OL}
A.~S. Solntsev and A.~A. Sukhorukov, \enquote{Combined frequency conversion and
  pulse compression in nonlinear tapered waveguides,} Opt. Lett. \textbf{37},
  446--448 (2012).

\bibitem{Almeida:2004-1081:NAT}
V.~R. Almeida, C.~A. Barrios, R.~R. Panepucci, and M.~Lipson,
  \enquote{All-optical control of light on a silicon chip,} Nature
  \textbf{431}, 1081--1084 (2004).

\bibitem{Vlasov:2008-242:NPHOT}
Y.~Vlasov, W.~M.~J. Green, and F.~Xia, \enquote{High-throughput silicon
  nanophotonic wavelength-insensitive switch for on-chip optical networks,}
  Nature Photonics \textbf{2}, 242--246 (2008).

\bibitem{deNobriga:2010-3925:OL}
C.~E. de~Nobriga, G.~D. Hobbs, W.~J. Wadsworth, J.~C. Knight, D.~V. Skryabin,
  A.~Samarelli, M.~Sorel, and R.~M. De~La~Rue, \enquote{Supermode dispersion
  and waveguide-to-slot mode transition in arrays of silicon-on-insulator
  waveguides,} Opt. Lett. \textbf{35}, 3925--3927 (2010).

\bibitem{Peleg:2009-163902:PRL}
O.~Peleg, M.~Segev, G.~Bartal, D.~N. Christodoulides, and N.~Moiseyev,
  \enquote{Nonlinear waves in subwavelength waveguide arrays: Evanescent bands
  and the "phoenix soliton",} Phys. Rev. Lett. \textbf{102}, 163902--4 (2009).

\bibitem{Benton:2009-5879:OE}
C.~J. Benton and D.~V. Skryabin, \enquote{Coupling induced anomalous group
  velocity dispersion in nonlinear arrays of silicon photonic wires,} Opt.
  Express \textbf{17}, 5879--5884 (2009).

\bibitem{Laegsgaard:2004-2473:OL}
J.~Laegsgaard, O.~Bang, and A.~Bjarklev, \enquote{Photonic crystal fiber design
  for broadband directional coupling,} Opt. Lett. \textbf{29}, 2473--2475
  (2004).

\bibitem{Garanovich:2007-475:OL}
I.~L. Garanovich and A.~A. Sukhorukov, \enquote{Nonlinear directional coupler
  for polychromatic light,} Opt. Lett. \textbf{32}, 475--477 (2007).

\bibitem{DellaValle:2010-673:OL}
G.~Della~Valle and S.~Longhi, \enquote{Subwavelength diffraction control and
  self-imaging in curved plasmonic waveguide arrays,} Opt. Lett. \textbf{35},
  673--675 (2010).

\bibitem{Garanovich:2012-1:PRP}
I.~L. Garanovich, S.~Longhi, A.~A. Sukhorukov, and Y.~S. Kivshar,
  \enquote{Light propagation and localization in modulated photonic lattices
  and waveguides,} Phys. Rep. \textbf{518}, 1--79 (2012).

\bibitem{Szameit:2009-271:NPHYS}
A.~Szameit, I.~L. Garanovich, M.~Heinrich, A.~A. Sukhorukov, F.~Dreisow,
  T.~Pertsch, S.~Nolte, A.~Tuennermann, and Y.~S. Kivshar,
  \enquote{Polychromatic dynamic localization in curved photonic lattices,}
  Nature Physics \textbf{5}, 271--275 (2009).

\bibitem{Dai:2007-2853:JOSB}
D.~X. Dai and Z.~Sheng, \enquote{Numerical analysis of silicon-on-insulator
  ridge nanowires by using a full-vectorial finite difference method mode
  solver,} J. Opt. Soc. Am. B \textbf{24}, 2853--2859 (2007).

\bibitem{Solntsev:2012-27441:OE}
A.~S. Solntsev, A.~A. Sukhorukov, D.~N. Neshev, and Y.~S. Kivshar,
  \enquote{Photon-pair generation in arrays of cubic nonlinear waveguides,}
  Opt. Express \textbf{20}, 27441--27446 (2012).

\end{thebibliography}
\end{document}